\documentclass{article}

\PassOptionsToPackage{numbers, compress}{natbib}
     \usepackage[final]{neurips_2019}

\usepackage[nonatbib]{}
\usepackage[utf8]{inputenc} % allow utf-8 input
\usepackage[T1]{fontenc}    % use 8-bit T1 fonts
\usepackage{hyperref}       % hyperlinks
\usepackage{url}            % simple URL typesetting
\usepackage{booktabs}       % professional-quality tables
\usepackage{amsfonts}       % blackboard math symbols
\usepackage{nicefrac}       % compact symbols for 1/2, etc.
\usepackage{microtype}      % microtypography
\usepackage[utf8]{inputenc} % allow utf-8 input
\usepackage[T1]{fontenc}    % use 8-bit T1 fonts
\usepackage{hyperref}       % hyperlinks
\usepackage{url}            % simple URL typesetting
\usepackage{booktabs}       % professional-quality tables
\usepackage{amsfonts}       % blackboard math symbols
\usepackage{nicefrac}       % compact symbols for 1/2, etc.
\usepackage{microtype}      % microtypography
\usepackage{multirow}
\usepackage{tabularx}
\usepackage{threeparttable}
\usepackage{booktabs}
\usepackage{bm}
\usepackage{amsmath}
\usepackage{stfloats}
\usepackage{textcomp}
\usepackage{gensymb}
\usepackage{graphicx}
\usepackage[T1]{fontenc}
\usepackage[utf8]{inputenc}
\usepackage{babel}
\usepackage[font=small,labelfont=bf,hypcap=false]{caption}
\usepackage[]{natbib}
\title{Adversarial cycle-consistent synthesis of cerebral microbleeds for data augmentation}

\author{%
  Khrystyna Faryna\\
  Diagnostic Image Analysis Group \\
  Radboud UMC, Nijmegen, 6525 GA \\
  \texttt{khrystyna.faryna@radboudumc.nl} \\
  \And
   Kevin Koschmieder \\
  Diagnostic Image Analysis Group \\
  Radboud UMC, 6525 GA Nijmegen \\
   \texttt{kevin.koschmieder@radboudumc.nl} \\
   \AND
  Marcella M. Paul \\
  Diagnostic Image Analysis Group \\
  Radboud UMC, Nijmegen, 6525 GA \\
   \texttt{marcella.paul@radboudumc.nl} \\
   \And
   Thomas van den Heuvel \\
  Dept. Of Medical Imaging \\
  Radboud UMC, Nijmegen, 6525 GA \\
   \texttt{thomas.vandenheuvel@radboudumc.nl} \\
   \And
   Anke van der Eerden \\
  Department of Radiology\\
  Erasmus MC, Rotterdam, 3015 CN \\ 
   \texttt{a.vandereerden@erasmusmc.nl} \\
   \And
   Rashindra Manniesing \\
   Diagnostic Image Analysis Group \\
  Radboud UMC, Nijmegen, 6525 GA \\
  \texttt{rashindra.manniesing@radboudumc.nl } \\
  \AND
  Bram van Ginneken \\
  Diagnostic Image Analysis Group \\
  Radboud UMC, Nijmegen, 6525 GA \\
   \texttt{bram.vanginneken@radboudumc.nl} \\
}

\begin{document}

\maketitle

\begin{abstract}
We propose a novel framework for controllable pathological image synthesis for data augmentation. Inspired by CycleGAN, we perform cycle-consistent image-to-image translation between two domains: healthy and pathological. Guided by a semantic mask, an adversarially trained generator synthesizes pathology on a healthy image in the specified location. We demonstrate our approach on an institutional dataset of cerebral microbleeds in traumatic brain injury patients. 
We utilize synthetic images generated with our method for data augmentation in cerebral microbleeds detection. Enriching the training dataset with synthetic images exhibits the potential to increase detection performance for cerebral microbleeds in traumatic brain injury patients. 

\end{abstract}

\section{Introduction}
Clinical outcome for patients with traumatic brain injury (TBI) is associated with cerebral microbleeds (CMBs) \citep{Werring2011CerebralMP}. CMBs are the result of leakages of small blood vessels, where hemosiderin deposits lead to focal dephasing of the MRI signal \citep{Roob991}. On MR-Susceptibility Weighted Imaging (SWI), CMBs appear as ``\textit{spherical hypointense lesions}''\cite{GREENBERG2009165}, with a diameter of less than ten millimeters. In TBI cases, CMBs can also have an elongated shape. Detection of CMBs, particularly their number and location, can provide useful information for the clinical prognosis of patients with TBI. However, due to their small size and visual similarity to blood vessels in a 2D view, manual annotation can be prohibitively time-consuming ($\geq1$ h/scan) for clinical diagnosis of patients with moderate to severe TBI \cite{Thomas}. While a number of methods for automatic detection of CMBs have been proposed \cite{LIU2019271,Standvoss2018CerebralMD}, a scarce availability of training samples with corresponding voxelwise annotations hampers the detection performance of these algorithms. In this study, we aim to generate synthetic TBI images (and the corresponding pathology mask) to augment the training dataset for a CMB detection system.

Synthesizing high-resolution images from random noise requires a vast number of training samples, thus, in this work we focus on synthesizing pathology on healthy images. We propose a data augmentation framework for synthesis of 3D medical images through pathology factorization and adversarial cycle-consistent learning. Our approach is an extension on CycleGAN \cite{zhu2017unpaired} by introducing pathology masks as additional input and ``\textit{abnormality mask}'' \cite{sun2018adversarial} loss, which encourages preservation of identity in regions outside of the pathology mask. We perform pathological-to-healthy and healthy-to-pathological image synthesis, factorizing pathology into a binary mask, to address the one-to-many problem which is axiomatic in the healthy-to-pathological synthesis task. We additionally modify the identity preservation pathway of the CycleGAN encouraging the lesions to be synthesized (in-painted) exclusively in the area specified by the annotation provided.
While the majority of methods in the literature focus on synthesis of unlabeled data \cite{wei2019generative}, patch in-painting \cite{Gupta19xray} or images belonging to restricted classes, our method is capable of synthesizing lesions of a size and location specified by a binary mask; it is not restricted to a particular imaging modality or abnormality \cite{Xavier}, neither does it require any additional surrounding tissue annotation \cite{shin2018medical}.  

\section{Materials and methods}

\paragraph{Data} The dataset used in this study consisted of 67 SWI scans from patients with moderate to severe TBI and 18 healthy subject SWI scans, collected at the Radboud University Medical Center, Nijmegen, The Netherlands. Scans have been acquired with the following parameters: TR:$27$ ms, TE:$20$ ms, voxel size: $0.98\times0.98\times1.00$ mm$^3$. A subset of 10 patient scans has been manually annotated by 6 medical observers, to be used as a comprehensive testing set. The remaining scans have been annotated by a single trained expert, subsequently split into training (46 pathological and 14 healthy scans) and validation (11 pathological and 4 healthy). First, we clip image intensities between $0$ and $99.5$ percentile and then rescale the range between 0 and 1. From each scan, we extract patches of $160\times146\times32$, capturing the whole axial plane of the brain, with a 50\% overlap in z direction. 
\paragraph{Lesion synthesis framework}

\begin{figure}[h]
\centering
\includegraphics[width=\textwidth]{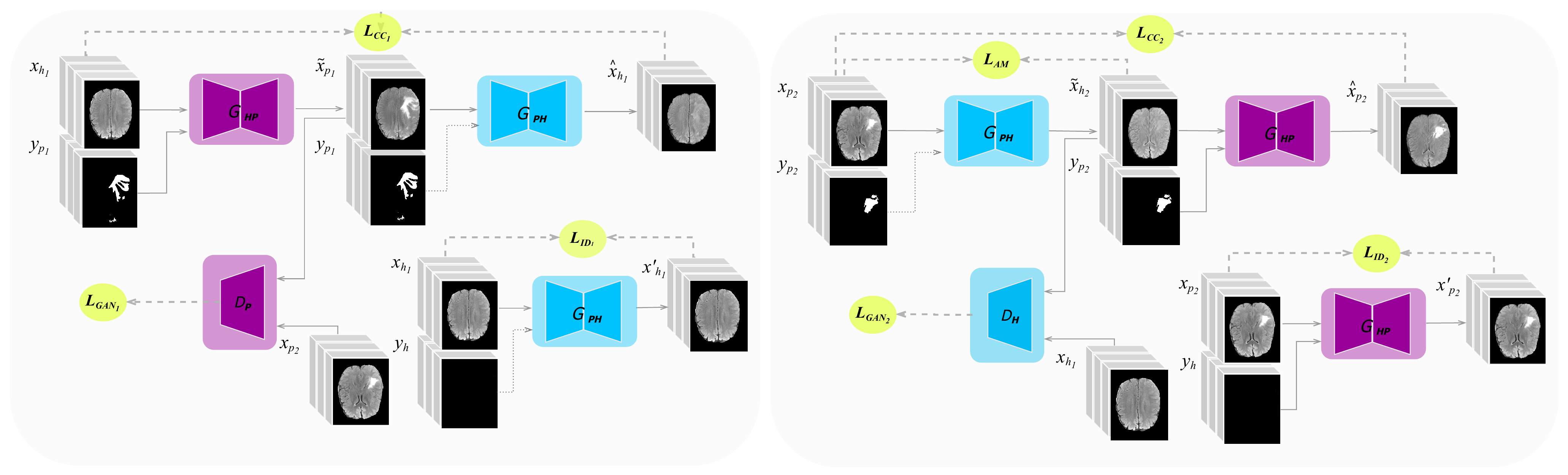}
\caption{The schematic of our proposed method: HPH cycle (right), PHP cycle (left). For visualization purposes, we use brain tumour samples from BraTS2018 \cite{Brats2018_1} dataset for this figure because their larger size compared to CMBs better displays the concept of semantic maps and their results.}
\label{fig:method}
\end{figure}

The schematic of the proposed approach along with training losses is shown in Fig.\ref{fig:method}. Inspired by CycleGAN, the configuration comprises  of two cycles: healthy-to-pathological-to-healthy (HPH) synthesis and pathological-to-healthy-to-pathological (PHP) synthesis. 
Here, we denote a healthy sample $x_{h_{i}}$ with an empty pathology mask $y_{h_{i}}$, where $x_{h_{i}}$ belongs to the healthy data distribution, $x_{h_{i}}$  $\sim$ \emph{H}. A pathological sample is denoted as $x_{p_{i}}$ with the corresponding pathology mask $y_{p_{i}}$, where $x_{p_{i}}$ belongs to the pathological data distribution, $x_{p_{i}}$ $\sim$ \emph{P}. Our objective is two-fold: given $x_{h_{i}}$ and a semi-random pathology mask $y_{p_{i}}$, generate a synthetic image $\tilde{x}_{p_{i}}$ such that $\tilde{x}_{p_{i}}$ $\sim$ \emph{P}, and given $x_{p_{i}}$ (and a suitable pathology mask $y_{p_{i}}$), generate a synthetic image $\tilde{x}_{h_{i}}$ such that $\tilde{x}_{h_{i}}$ $\sim$ \emph{H}. The index \textit{i} specifies a particular subject.

In the HPH cycle, the healthy-to-pathological generator $G_{HP}$ receives as input the healthy image $x_{h_{1}}$ and a semi-random pathology mask $y_{p_{1}}$. The objective of $G_{HP}$ is to synthesize a pseudo-pathological image $\tilde{x}_{p_{1}}$, such that the pathology would be located in the areas provided by the mask $y_{p_{1}}$, while the rest of the image should remain unchanged. The goal of the discriminator $D_{P}$ is to distinguish between real ($x_{p_{2}}$) and fake ($\tilde{x}_{p_{1}}$) pathological samples. The obtained pseudo-pathological image $\tilde{x}_{p_{1}}$ is then concatenated with its pathology mask $y_{p_{1}}$ and passed to the pathological-to-healthy generator $G_{PH}$. The task of $G_{PH}$ is to reconstruct the input image $\hat{x}_{p_{2}}$. Additionally, to encourage preservation of identity and assure that $G_{PH}$ does not modify a healthy image the $x_{h_{1}}$ is concatenated with $y_{h}$ and passed through $G_{PH}$. The losses used in this cycle are: cycle-consistency ($L_{CC}$), identity ($L_{ID}$), and adversarial loss ($L_{GAN}$). The latter is the Wasserstein loss with gradient penalty\cite{Gulrajani2017}.  The reverse is happening in PHP cycle, with only one exception--the use of abnormality mask loss ($L_{AM}$)\cite{sun2018adversarial} to encourage preservation of brain structures outside of the pathology region.

The architectures of $G_{PH}, G_{HP}, D_{P}$ and $D_{H}$ are inspired by \cite{XIAMIA2020}, however we expanded on the method by switching to a 3D approach, exchanging all 2D layers with their 3D counterparts. The model was trained using the Adam\cite{Kingma2015AdamAM} optimizer with $\beta_{1}$=0.5, $\beta_{2}$=0.99, learning rate of 0.0001 and a batch size of 4. 
\paragraph{Lesion detection}
The detection model used in this study consists of two stages: candidate proposal with Fast Radial Symmetry Transform \cite{loy2003fast} and a classification CNN. The experiment is composed of three pipelines, the detection network is trained with only real, synthetic and combined real and synthetic data. We perform additional experiment comparing performance using also classical data augmentation (cDA), in particular flipping in the axial plane, shearing, scaling, rotation, and intensity (scaling, shifting) augmentations.

\section{Results and discussion}

\begin{figure}[h]
\centering
\includegraphics[width=\textwidth]{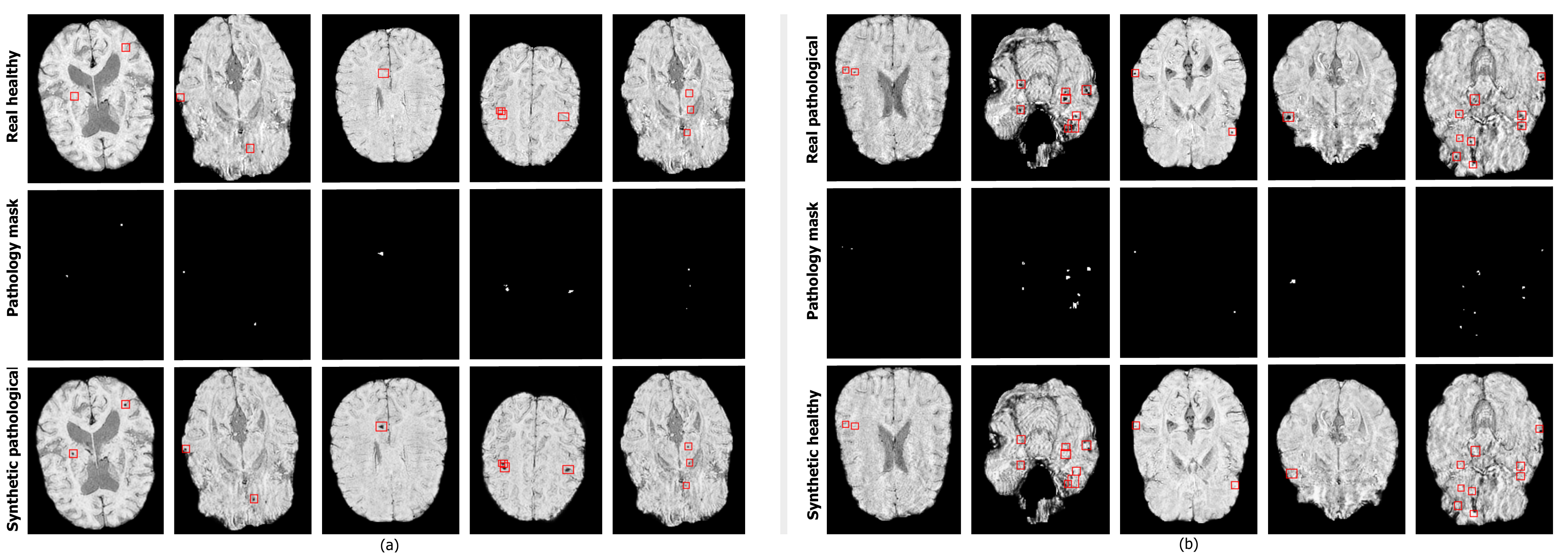}
\captionsetup{type=figure}
\caption{Synthetic SWI images generated with the proposed method: (a)--generation of synthetic pathological images, generation of synthetic healthy images. The red bounding boxes indicate the place where the lesion is generated/in-painted. }
\label{fig:brains}
\end{figure}

Examples of synthetic healthy and pathological images produced with the proposed method are shown in Fig. \ref{fig:brains}. Detection performance of models in different settings was evaluated using a Free-response Receiver Operating Characteristic (FROC) curve, bootstrapping was used to compute 95\% confidence intervals (CI). Results of the experiments with synthetic data augmentation are shown in Fig. 3 and Table 1. 

\begin{minipage}[b]{0.35\textwidth}
    \centering
    %\captionsetup{type=tabel}
\captionof{table}{Sensitivity of models trained on different data distributions compared at a fixed, clinically relevant rate of 10 false positives (FPs) per TBI patient.}
\scriptsize
\begin{tabular}[b]{lrc}
    \toprule    %[0.3pt] 
Training data & & Sensitivity (\%) \\
    \midrule    %[0.3pt]
    \toprule    %[0.3pt] 
\noalign{\smallskip}
Real & & 77.8  \\
Synthetic & & 71.8 \\
Real+Synthetic & & \textbf{78.9} \\
\midrule
Real & +cDA & 80.5  \\
Synthetic & +cDA & 80.1  \\
Real+Synthetic & +cDA & \textbf{82.4}  \\

    \bottomrule [0.3pt]
\end{tabular}
\end{minipage}
\hfill
\begin{minipage}[b]{0.6\textwidth}
   % \centering
     \includegraphics[width=\textwidth]{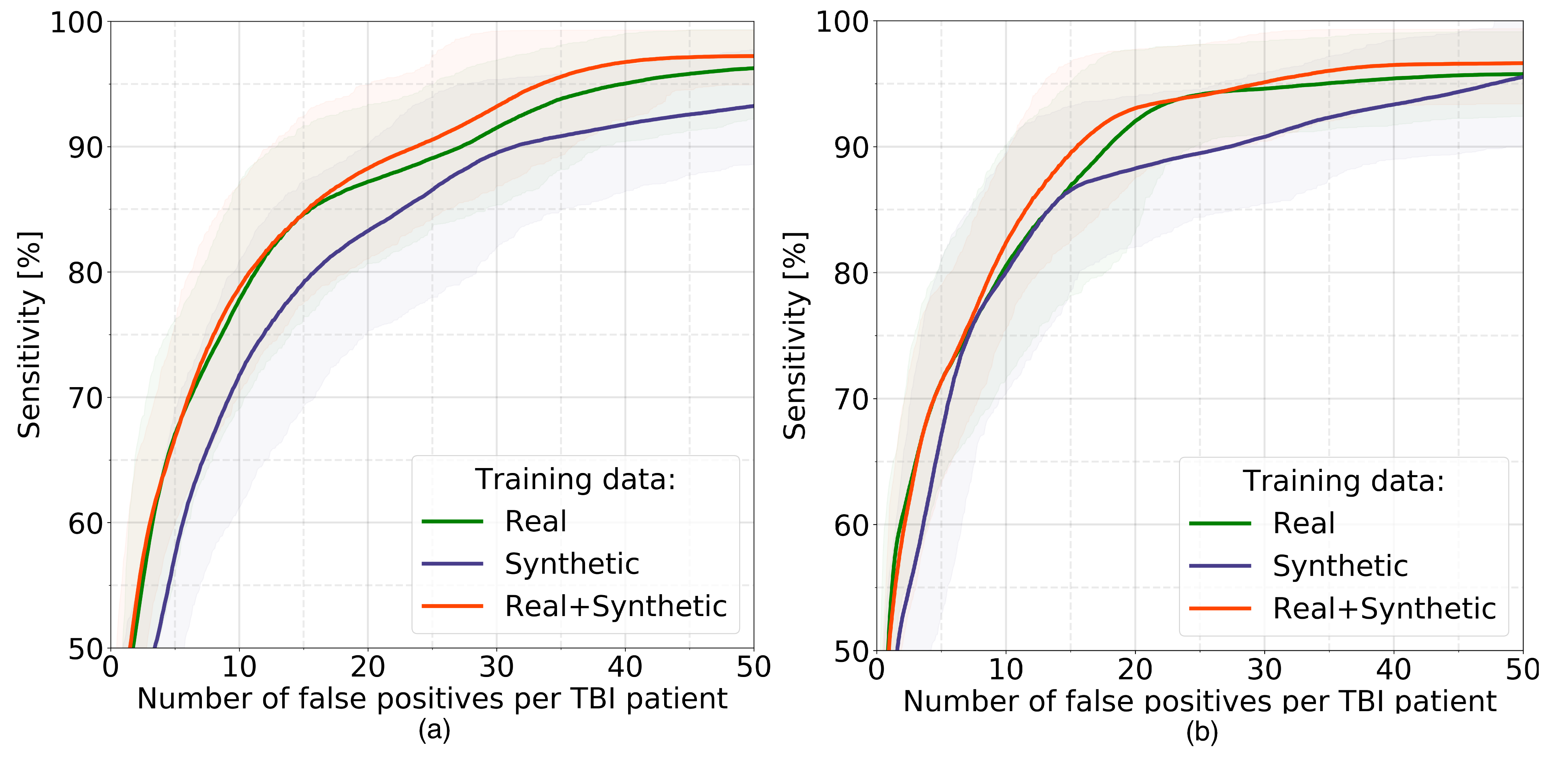}
\captionof{figure}{FROC performance with 95\% CI on an independent test set of models trained with different (real, synthetic, real+synthetic) data: (a)--without cDA, (b)--with cDA. }
    \end{minipage}

The models trained on synthetic data are capable of producing meaningful predictions when tested on real samples, demonstrating comparatively lower performance: 90\% sensitivity is reached at 27.4 and 31.4 FPs/patient with and without cDA, respectively. The models trained with real data achieve a sensitivity of 90\% at 17.9 and 27.3 FPs/patient with and without cDA. Combining the real samples with synthetic ones with 1:1 ratio, we observe a marginal yet consistent improvement in performance with respect to the models trained on real data only: the models reach 90\% sensitivity at 23.7 and 15.5 FPs/patient. We note here that intersections between CIs of the evaluated methods are relatively high, nevertheless we envision the prospective value of using the trained system for generating CMB lesions on external SWI data, potentially reducing the burden of additional manual annotation while dealing with data coming from different institutions. 
 
In conclusion, we proposed a method for synthesis of pathology on healthy medical images guided by a pathology mask and demonstrated its use in data augmentation for CMB detection in TBI patients.
\section{Broader impact}

Detection of CMBs in TBI patients is a prime example of a medical application that is heavily afflicted by data scarcity and variations between MR scanners. The proposed approach is aimed at reducing the burden of manual annotation while shifting between various vendors and imaging protocols.

\section{Acknowledgements}
Khrystyna Faryna was supported by Erasmus Mundus Joint Master Degree in Medical Imaging and Applications (MAIA) grant.
We wish to thank Prof. Xavier Llado for his valuable feedback.
We would like to express our gratitude to T.M.J.C. Andriessen and P. Vos for collecting the data, and to T.M.J.C. Andriessen, T. Vande Vyvere, L. van den Hauwe, B. Geurts and B.M. Goraj for their work in annotating the Cerebral Microbleeds.

\bibliography{bibl}

\end{document}